\documentclass[%
reprint,
superscriptaddress,
 amsmath,amssymb,
 aps,
 twocolumn,
floatfix,
]{revtex4-2}

\usepackage[utf8]{inputenc}
\usepackage{graphicx}
\usepackage{dcolumn}
\usepackage{bm}
\usepackage{hyperref}
\usepackage{booktabs}
\usepackage{siunitx}
\usepackage{caption}
\usepackage{subcaption}
\usepackage{xcolor}
\usepackage[normalem]{ulem}
\usepackage[bottom]{footmisc}		
\usepackage{comment}
\begin{document}
\preprint{}
\title{
A fast tracking code for evaluating collective effects in linear accelerators
}

\author{F. Bosco}\thanks{Corresponding author. Email: fabio.bosco@uniroma1.it}
\affiliation{Sapienza University, Dept. of Basic and Applied Sciences for Engineering, via A. Scarpa 14, 00161 Roma, Italy}
\affiliation{INFN–Sezione di Roma, P.le Aldo Moro 2, Roma 00185, Italy}
\affiliation{University of California, Dept. of Physics and Astronomy, 405 Hilgard Ave, Los Angeles, CA 90095, USA}

\author{O. Camacho}
\affiliation{University of California, Dept. of Physics and Astronomy, 405 Hilgard Ave, Los Angeles, CA 90095, USA}

\author{M. Carillo}
\affiliation{Sapienza University, Dept. of Basic and Applied Sciences for Engineering, via A. Scarpa 14, 00161 Roma, Italy}
\affiliation{INFN–Sezione di Roma, P.le Aldo Moro 2, Roma 00185, Italy}

\author{E. Chiadroni}
\affiliation{INFN–Laboratori Nazionali di Frascati, Via E. Fermi 40, Frascati (Roma) 00044, Italy}
\affiliation{Sapienza University, Dept. of Basic and Applied Sciences for Engineering, via A. Scarpa 14, 00161 Roma, Italy}

\author{L. Faillace}
\affiliation{INFN–Laboratori Nazionali di Frascati, Via E. Fermi 40, Frascati (Roma) 00044, Italy}
\affiliation{Sapienza University, Dept. of Basic and Applied Sciences for Engineering, via A. Scarpa 14, 00161 Roma, Italy}

\author{A. Fukasawa}
\affiliation{University of California, Dept. of Physics and Astronomy, 405 Hilgard Ave, Los Angeles, CA 90095, USA}

\author{A. Giribono}
\affiliation{INFN–Laboratori Nazionali di Frascati, Via E. Fermi 40, Frascati (Roma) 00044, Italy}

\author{L. Giuliano}
\affiliation{Sapienza University, Dept. of Basic and Applied Sciences for Engineering, via A. Scarpa 14, 00161 Roma, Italy}
\affiliation{INFN–Sezione di Roma, P.le Aldo Moro 2, Roma 00185, Italy}

\author{N. Majernik}
\affiliation{University of California, Dept. of Physics and Astronomy, 405 Hilgard Ave, Los Angeles, CA 90095, USA}

\author{A. Mostacci}
\affiliation{Sapienza University, Dept. of Basic and Applied Sciences for Engineering, via A. Scarpa 14, 00161 Roma, Italy}
\affiliation{INFN–Sezione di Roma, P.le Aldo Moro 2, Roma 00185, Italy}

\author{L. Palumbo}
\affiliation{Sapienza University, Dept. of Basic and Applied Sciences for Engineering, via A. Scarpa 14, 00161 Roma, Italy}
\affiliation{INFN–Sezione di Roma, P.le Aldo Moro 2, Roma 00185, Italy}

\author{J.B. Rosenzweig}
\affiliation{University of California, Dept. of Physics and Astronomy, 405 Hilgard Ave, Los Angeles, CA 90095, USA}
\affiliation{Sapienza University, Dept. of Basic and Applied Sciences for Engineering, via A. Scarpa 14, 00161 Roma, Italy}

\author{B. Spataro}
\affiliation{INFN–Laboratori Nazionali di Frascati, Via E. Fermi 40, Frascati (Roma) 00044, Italy}

\author{C. Vaccarezza}
\affiliation{INFN–Laboratori Nazionali di Frascati, Via E. Fermi 40, Frascati (Roma) 00044, Italy}

\author{M. Migliorati}
\affiliation{Sapienza University, Dept. of Basic and Applied Sciences for Engineering, via A. Scarpa 14, 00161 Roma, Italy}
\affiliation{INFN–Sezione di Roma, P.le Aldo Moro 2, Roma 00185, Italy}
%

\begin{abstract}
The demands on performance of advanced linear accelerator based facilities strongly depend on the quality of the particle beams produced by such machines. Indeed, state-of-the-art applications in photon production and high-energy physics colliders require to use very high brightness electron beams, implying the coexistence of high peak currents and small transverse emittances. 
In such systems, the nominal phase-space density may be diluted by the presence of self-induced electromagnetic fields, causing interaction among charged particles through space charge forces and the excitation of wakefields. The two sources of collective effects may both be present in significant levels, and be coupled by the strong externally applied transverse and longitudinal fields present in modern high gradient linear accelerators.
Therefore, beam dynamics studies investigating all relevant effects, applied and collective, are necessary to predict the operational limitations of a given instrument. Such modeling, involving a large number of computational particles, can require significant numerical resources. 
Motivated by the urgency and attendant challenges in the above-described scenario, in this paper we thus present a fast tracking code which permits accurate evaluation of wakefield effects in rf linacs, while also including a simple, robust model for space-charge forces to streamline the computations.
The features of such a tool are discussed in detail in this paper and comparisons with more time-intensive commonly used tracking codes or analytical models are utilized to validate the approach we introduce.
In addition, the applications motivating the development of this code define unique and challenging scenarios from the perspective of beam physics.
Specifically, the fast simulation framework developed in this paper is particularly aimed at the challenges of describing the dynamics of intense electron beams injected at low energy - implying the existence of strong space-charge forces. Further, these intense beams propagate  in high-gradient accelerating structures which provide both strong rf focusing effects as well as giving rise to strong wakefield interactions. The enhancement of the wakefield coupling is due to the small iris dimensions present in optimized high  impedance, high gradient structure.
\end{abstract}

\maketitle

\section{Introduction}

High brightness electron beams, such as those produced by radio frequency (RF) photo-injectors \cite{CARLSTEN1989Newphotoelectric}, enable numerous state-of-the-art applications such as free electron lasers (FELs) \cite{pellegrini2016thephysics,emma2010Firstlasing}, inverse Compton scattering (ICS) sources \cite{serafini2017analytical,faillace2019status}, laser or particle driven wakefield acceleration (LWFA/PWFA) \cite{FERRARIO2018eupraxia} and linear colliders \cite{shiltsev2021modern}.
The growing interest for such attractive applications has motivated to exploit advanced accelerator and beam physics concepts in order to enhance the performance of future machines.
In particular, special efforts are addressed to produce high brightness electrons and to accelerate such beams within high gradient linacs.

Improvements of the beam brightness are made possible by novel rf photo-injectors either in the traditional standing wave (SW) or in the hybrid standing wave/traveling wave (SW/TW) \cite{spataro2011rf,rosenzweig2011design} configuration. Recent designs have proposed a high field SW cryo-cooled gun with optimized cell shape \cite{rosenzweig2018ultra,rosenzweig2019next,robles2021versatile} and a hybrid gun \cite{faillace2021beam,Faillace2022highfield} both working in C-band at $\SI{5.712}{GHz}$.
In addition, operation at cryogenic temperatures reduces the breakdown rate probability in rf cavities \cite{cahill2018_1,cahill2018_2} which allows to achieve higher accelerating gradients also in the main linac.
Moreover, a recent design concept proposed by SLAC, known as distributed coupling, allows a further enhancement of the accelerating gradient \cite{tantawi2018distributed,tantawi2020design}.
Such a technique exploits waveguide manifolds to feed the rf power individually in each cell of periodic structures. As a result, the iris dimensions are no longer constrained to coupling requirements and the individual cell shape can be optimized to maximize the acceleration process.
This clever concept has been embraced for the realization of cutting edge facilities based on C-band linacs such as ultra-compact X-ray FELs \cite{ucxfel}, low bandwidth and high flux ICS machines \cite{faillace2021beam,Faillace2022highfield} and linear colliders \cite{Ccube,whitepaperSnowmass}.






The above features combined together define a unique scenario for the beam dynamics of high brightness electrons for the following reasons.
\begin{itemize}
    \item The high accelerating gradient and the rich content of space harmonics arising from the optimized cavity topology have a strong impact on the transverse optics \cite{hartman1993ponderomotive,rosenzweig1994transverse}. As we will discuss in the next section, such features make possible to provide radial focusing in booster linacs following the gun without need for solenoid magnets.
    \item Despite the strong acceleration rapidly increases the beam momentum, the low energy regime, which is characteristic of rf photo-injectors followed by a booster linac, plays a crucial role. Indeed, electrons are injected in the main accelerator with kinetic energies in the range $\sim$ 4-6 MeV which implies a non-negligible residual sensitivity to space charge effects. Such forces act against the external focusing whose strength has to be chosen accordingly to achieve an equilibrium condition known as the invariant envelope criterion \cite{serafini1997envelope}.
    \item In addition, the optimized cell shape employed in distributed coupling linacs exhibits a small iris radius, $a = \SI{2}{mm}$ at C-band (\emph{i.e} $a/\lambda\approx0.038$), in order to enhance the effective shunt impedance.
However, the small iris dimension is also accompanied by stronger wakefield effects \cite{chao1993physics,PalumboVaccaroZobov} which perturb the motion of the trailing particles leading to energy spread, emittance dilution and transverse instabilities such as the \emph{beam break-up} (BBU) \cite{panofsky1968asymptotic}.
\end{itemize}

Such a challenging environment has motivated us to investigate the effects of wakefield interaction in linear accelerators characterized by the peculiar properties listed above.
Several types of tracking codes are in use in the field of accelerator physics \cite{codesAccPhys}. Their purpose is to describe the beam dynamics in different operating conditions which typically establish the physical processes that must be included in the model and those that can be neglected.
For instance, some codes \cite{elegant} describe wakefield effects but they do not include self-consistent models for the longitudinal or transverse space charge, which is crucial in space charge-dominated beams. Conversely, other codes \cite{ref_GPT,parmela} include space charge forces neglecting the wakefields.
Including both types of collective effects increases the complexity of the models as well as the computational times due to the high number of operations and, thus, it is not a common feature. The code ASTRA \cite{ASTRA}, for instance, describes both mechanisms by using particle-in-cell methods for space-charge forces and convolution integrals with the longitudinal beam slices for wakefields.

In this paper we investigate alternative approaches based on simpler models that allow to account for wakefield effects in presence of space charge forces.
Such models are combined in a dedicated tracking code named MILES (\emph{Modeling Instabilities in Linacs with Ellipsoidal Space charge}) providing a light and flexible tool whose preliminary version was introduced in \cite{bosco2021modeling}.
Compared to codes employing standard approaches, MILES allows for fast simulation times (typically a factor 1/20-1/10) at the expenses of an acceptable reduction of the accuracy.
However, it should be stressed that the simplified models which are valid for linacs are not sufficient for the non-relativistic, near-cathode region.
Therefore, our work is not aimed at producing start-to-end simulations, or to provide a complete alternative for codes that include the emission process in photo-injectors.
Our analyses focus instead on the dynamics starting from the main linac with particular emphasis on wakefield effects which can act over very long integrated distances. Therefore, in our tracking code, we typically begin by importing a particle distribution that provides the initial conditions at the exit of the low-energy injection region.

Throughout the paper we describe the models adopted by MILES in order to include the main aspects involved in the beam dynamics.
The work is organized as follows. In Section II we introduce the basic equations describing the transport mechanisms in the linacs of our interest in absence of collective effects.
Section III and IV introduce the models adopted by MILES for the description of space charge forces and wakefield effects respectively.
In Section V the subject of emittance dilution induced by the transverse wakefields excited in misaligned linac sections is discussed, while in Section VI we investigate possible correction techniques exploiting beam based alignment concepts.
Finally, in Section VII we mention some additional applications for which our code can be employed.

\section{Single particle dynamics in rf linacs}

Electron beams propagating in axi-symmetric linear accelerators constitute a dynamical system whose evolution can be suitably described by tracking a set of dynamical variables, as reviewed here. 
The model we adopt describes the beam as an ensemble of charged macro-particles which form an axi-symmetric distribution, and are characterized by use of the following quantities:

\begin{itemize}
\item horizontal deviation from the nominal linac axis, $x$
\item horizontal divergence with respect to the nominal linac axis, $x^\prime$
(with the prime symbol denoting derivatives with respect to the longitudinal position $z$)
\item longitudinal deviation from the beam's centroid, $\zeta = z- \beta_0 c t = z - \langle z \rangle$
where the brackets imply an average on the distribution, $\beta_0 c$ is the mean beam velocity and $t$ is the design time of arrival
\item total particle energy, $E = \gamma mc^2$ (with $\gamma = 1/\sqrt{1 - \beta^2}$, $\beta$ being the velocity of the macro-particle in speed of light units)
\end{itemize}

As the beam propagates along the axis of the linac such quantities transform under the action of both applied and self-induced electromagnetic fields.
As we will explain later, the rotational symmetry is perturbed by the transverse wakefields which introduce non-axysimmetric motion. Thus, the validity of such an assumption is investigated further in the sections dedicated to collective effects.
In this section we discuss the effects of applied fields on charged particles and show the relevant transfer maps which describe the evolution of the dynamical variables for our cases of interest.

Focusing in both transverse dimensions of charged beams in high energy accelerators is conventionally provided by use of alternating gradient quadrupole magnets arranged in a periodic lattice \cite{COURANT19581}.
However, the low energy dynamics of space charge dominated beams, which is characteristic of rf photo-injectors followed by a booster linac, requires a different approach. Indeed, the rms spot size is typically matched to an equilibrium-like propagation mode known as the \emph{invariant envelope} which allows optimization of the emittance compensation process \cite{serafini1997envelope,WavebreakSGA}.
In this scenario the radial focusing can be provided by the use of solenoid magnets, the non-synchronous space harmonics in accelerating structures, or their superposition. 

Intense charged particle beams accelerated in multi-cell linacs inherently experience two types of radial forces which affect the dynamics in high gradient structures.
The first mechanism is introduced by the non-synchronous space harmonics of the RF wave in the periodic linac cavities. Due to the dependence of the averaged force on the accelerating gradient, this effect is known as \emph{second order} (or ponderomotive) rf  focusing \cite{hartman1993ponderomotive}.
The linear transfer map describing the evolution of \emph{secular} trajectories in a rf cell of length $L_c$ and smoothed accelerating gradient $\gamma^\prime = \langle eE_z\rangle/mc^2$ is given by \cite{rosenzweig1994transverse}

\begin{widetext}
\begin{equation}
    \begin{pmatrix}
        x \\
        x^\prime
    \end{pmatrix}
    \mapsto
    \begin{pmatrix}
        \cos{\left(\nu\ln\frac{\gamma + \gamma^\prime L_c}{\gamma}\right)} & \frac{\gamma}{\nu\gamma^{\prime}}\sin{\left(\nu\ln\frac{\gamma + \gamma^\prime L_c}{\gamma}\right)}\\
        -\frac{\nu\gamma^{\prime}}{\gamma + \gamma^\prime L_c}\sin{\left(\nu\ln\frac{\gamma + \gamma^\prime L_c}{\gamma}\right)} & \frac{\gamma}{\gamma + \gamma^\prime L_c}\cos{\left(\nu\ln\frac{\gamma + \gamma^\prime L_c}{\gamma}\right)}
    \end{pmatrix}
    \begin{pmatrix}
        x \\
        x^\prime
    \end{pmatrix}
    \label{eq:jrls_matrix}
\end{equation}
\end{widetext}
where $\nu = \sqrt{\eta/8}/\vert\cos{\Delta\phi}\vert$, $\Delta\phi$ is the phase deviation from the rf crest and $\eta(\Delta\phi)$ quantifies the space harmonic content.
Furthermore, the matrix in Eq. (\ref{eq:jrls_matrix}) allows inclusion of the superposition of a solenoidal magnetic field of strength $B_z$ by a simple substitution $\eta\mapsto\eta + 2 b^2$ where $b = cB_z/\langle E_z\rangle$.
As an additional focusing effect, particles entering or leaving an accelerating section experience a transition from a field-free to a non-zero field region, or vice versa. The change in the longitudinal field gives rise to a radial field component which is responsible for a transverse kick.
Such an effect, known also as \emph{first order} rf focusing, can be described in terms of a thin lens matrix (see \cite{rosenzweig1994transverse})
\begin{equation}
    \begin{pmatrix}
        x \\
        x^\prime
    \end{pmatrix}
    \mapsto
    \begin{pmatrix}
        1 & 0\\
        \mp\frac{\gamma^{\prime}}{2\gamma} & 1
    \end{pmatrix}
    \begin{pmatrix}
        x \\
        x^\prime
    \end{pmatrix}.
    \label{eq:lens rf}
\end{equation}
This effect is extremely important at the entrance (\emph{i.e.} minus sign) of the first accelerating section where the beam possesses its lowest energy.





For completeness, we also show the transfer map for the longitudinal variables inside an accelerating cell which simply follows the conventional equations of rf linacs \cite{wangler}.
Such processes account for particles propagating forwards in the machine while gaining energy and is described by the following transformations
\begin{equation}
    \begin{pmatrix}
        z \\
        E
    \end{pmatrix}
    \mapsto
    \begin{pmatrix}
        z \\
        E
    \end{pmatrix} + L_c
    \begin{pmatrix}
        1 + \Delta\zeta/L_c \\
        eE_\text{acc}\cos\Delta\phi
    \end{pmatrix}
\end{equation}
where $\Delta\zeta$ is the longitudinal slippage with respect to the beam centroid and, at zero-th order, it is proportional to the corresponding differences in time of flight.

\begin{figure}
    \centering
    \begin{subfigure}{.96\columnwidth}
         \centering
         \includegraphics[width=\textwidth]{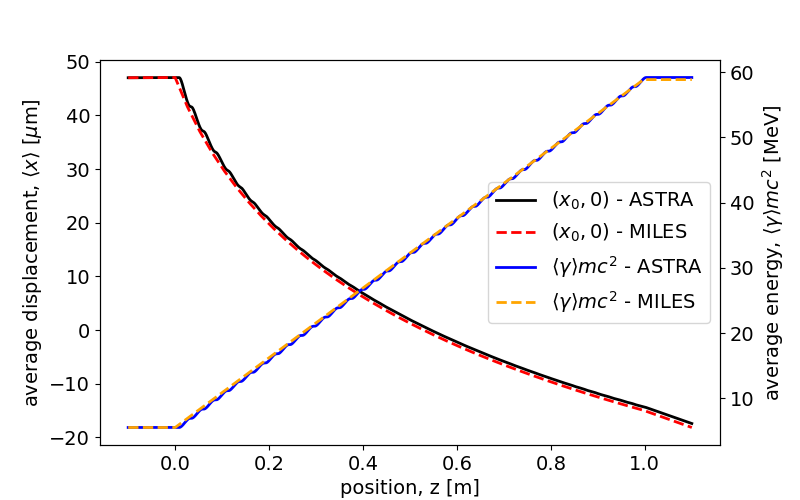}
         \caption{Injection at $\SI{5}{MeV}$ kinetic energy.}
         \label{avg_x_SCoff10_5MeV}
     \end{subfigure}
     \hfill
     \begin{subfigure}{.96\columnwidth}
         \centering
         \includegraphics[width=\textwidth]{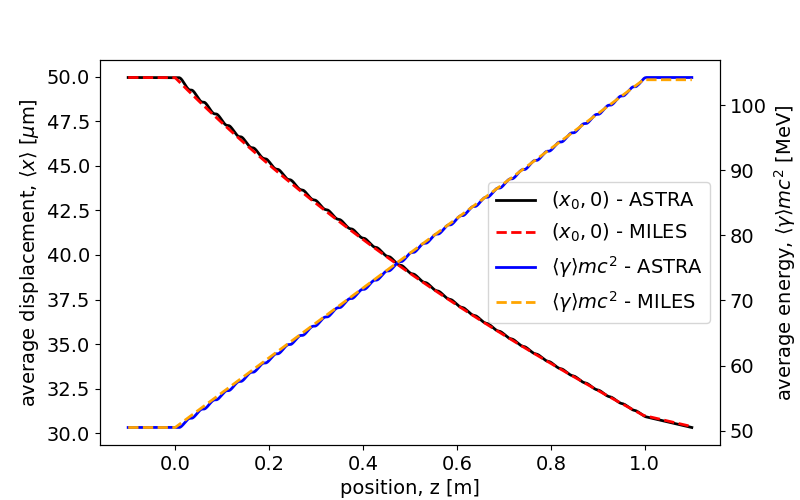}
         \caption{Injection at $\SI{50}{MeV}$ kinetic energy.}
         \label{avg_x_SCoff10_50MeV}
     \end{subfigure}
    \caption{Center of mass trajectory and mean energy for a beam injected in an accelerating section with initial conditions $(x_0,x^\prime_0)=(\SI{50}{\mu m},0)$. Solid curves represent results from ASTRA while dashed curves are obtained with MILES.}
    \label{fig:avg_x_SCoff}
\end{figure}

The rest of the section is dedicated to show examples of tracking in absence of collective effects. In this demonstration, the models introduced for the transverse optics are employed in MILES and the results compared with those from the well-known code ASTRA \cite{ASTRA}.
Since Eq. (\ref{eq:jrls_matrix}) and (\ref{eq:lens rf}) are derived in the approximation $\gamma\gg 1$, we investigate two reference cases for which the initial kinetic energy of the beam is 5 MeV and 50 MeV respectively.
In either cases we assume monochromatic rectangular beams with uncorrelated bi-gaussian transverse distribution with $\sigma_x=\SI{0.5}{mm}$ and $\epsilon_n=\SI{1}{\mu m\cdot m}$. Such beams are injected into a distributed coupling $\sim\SI{1}{m}$-long C-band linac section with $\langle eE_z\rangle = \SI{50}{MeV/m}$ accelerating gradient and second order focusing strength $\eta = 1.12 - 0.5\cos(2\Delta\phi)$ \cite{robles2021versatile}.
As an additional comment, the beam moments as well as the linac parameters in our benchmark tests usually match those for the ICS source in \cite{Faillace2022highfield} unless stated differently.
Figure \ref{fig:avg_x_SCoff} shows the center-of-mass trajectory and the energy for a beam injected with a transverse offset of $x_0 = \SI{50}{\mu m}$.
Solid lines are obtained with ASTRA while dashed lines are obtained with MILES showing the agreement is very good even in the low energy scenario.



\begin{figure}[tbh!]
    \centering
    \begin{subfigure}{.96\columnwidth}
         \centering
         \includegraphics[width=\textwidth]{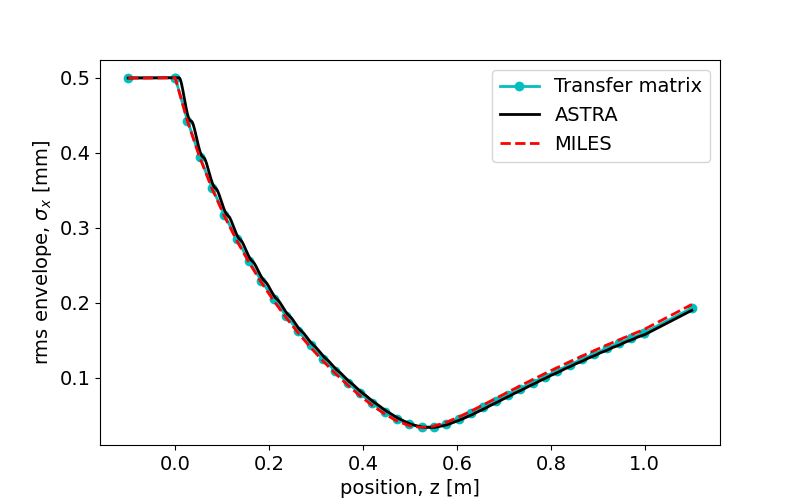}
         \caption{Injection at $\SI{5}{MeV}$ kinetic energy.}
         \label{std_x_SCoff_5MeV}
     \end{subfigure}
     \hfill
     \begin{subfigure}{.96\columnwidth}
         \centering
         \includegraphics[width=\textwidth]{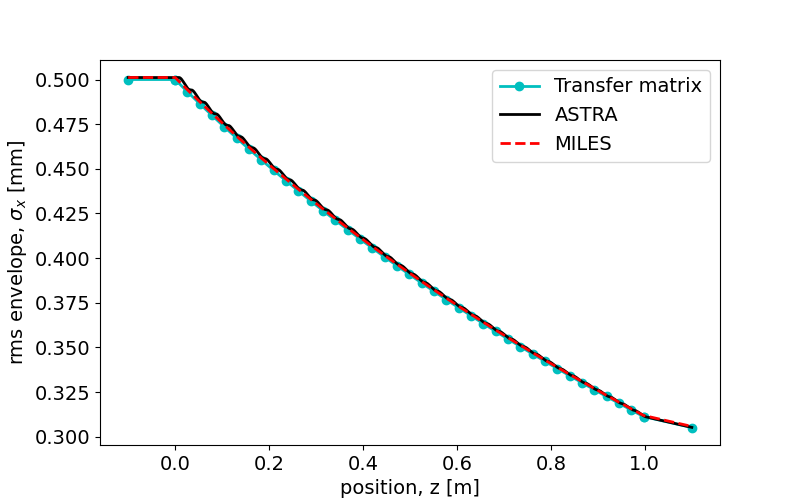}
         \caption{Injection at $\SI{50}{MeV}$ kinetic energy.}
         \label{std_x_SCoff_50MeV}
     \end{subfigure}
    \caption{Rms transverse envelope for a space charge-free electron beam propagating inside an accelerating section. The dots represent the analytic result from the transfer map while the black and red curves are obtained with ASTRA and MILES, respectively.}
    \label{fig:std_x_SCoff}
\end{figure}

Let us now focus on the evolution of the rms transverse envelope in the same accelerating structure.
Neglecting again the space charge forces, the evolution of the transverse phase space can be easily described by
\begin{subequations}
    \begin{align}
    &
    \begin{pmatrix}
        x \\
        x^\prime
    \end{pmatrix}
    \mapsto
    R
    \begin{pmatrix}
        x \\
        x^\prime
    \end{pmatrix}\\
    &
    \begin{pmatrix}
        \sigma_x^2 & \sigma_{xx^\prime} \\
        \sigma_{xx^\prime} & \sigma_{x^\prime}^2
    \end{pmatrix}
    \mapsto
    R
    \begin{pmatrix}
        \sigma_x^2 & \sigma_{xx^\prime} \\
        \sigma_{xx^\prime} & \sigma_{x^\prime}^2
    \end{pmatrix}
    R^T
    \label{eq:rms_param linear evolution}
    \end{align}
\end{subequations}
where $R$, in the linear approximation, is a matrix representing any of the transformations affecting the beam  \cite{wolski}.
Knowledge of the transfer matrix in Eq. (\ref{eq:jrls_matrix}) and (\ref{eq:lens rf}) allows to calculate analytically the evolution of the rms envelope for arbitrary initial conditions. In this example we assume that the injection occurs on-axis and that the initial rms parameters are $\sigma_x = \SI{0.5}{mm}$, $\sigma_{xx^\prime} = 0$ while $\sigma_{x^\prime}$ can be derived from the $\epsilon_n = \SI{1}{\mu m \cdot rad}$ normalized emittance.
In Fig. \ref{fig:std_x_SCoff} the cyan dots represent the evolution of the rms spotsize calculated with Eq. (\ref{eq:rms_param linear evolution}). Further, the solid black line is obtained by tracking the beam with ASTRA while the dashed red line is obtained by MILES, which demonstrates the expected behavior.


\section{Modeling space charge forces}

The low energy electron dynamics are strongly affected by Coulomb repulsion, which dominates especially in region of rf photoinjectors before the end of the first post-acceleration linac. Such forces thus crucially affect also the optics properties in the first linac sections, where particles are typically injected with energies in the 4-6 MeV range.
Indeed, the matching of laminar beams to a linac requires adjustment of the strength of both the applied focusing and the accelerating gradient as a function of the beam current in order to achieve the desired generalized equilibrium-like (\textit{i.e.} invariant envelope) condition \cite{serafini1997envelope}.
Consequently, space charge forces have a major impact on the low-energy evolution of the rms beam envelope as well as the rms emittance, and they must be taken into account.
As an example, the nominal supplied focusing which optimally controls the envelope evolution will be anomalously strong in absence of space charge forces causing over focusing of the trajectories with consequent non-laminar motion.

As the strength of the self-induced forces acting on the beam envelope depends on its size and distribution, a self-consistent dynamical model is required. 
The solution for such a problem is typically calculated exploiting computer codes except for some special cases which admit exact solutions.
Among these, the results for uniform ellipsoidal beams play a special role.
Of particular relevance to this discussion is the analysis of the fields of beams with uniform ellipsoidal transverse distributions, a study well-known in the literature \cite{KV_dist}. As we will discuss in this section, the self-induced force components for uniform three-dimensional ellipsoids of arbitrary aspect ratios exhibit a 
linear dependence on distance from centroid. Ideally, this linearity (\textit{e.g.} $F_x\propto x$) allows one to preserve the rms-emittances. This attractive property has motivated studies of reliable methods to produce beams with such distribution exploiting the transverse shape of the laser pulse at the photo-cathode \cite{luiten2004,pietro2008}. In the most detailed example given in this paper, the distribution of the beam very quickly becomes a nearly perfect ellipsoid in shape.


\subsection{Ellipsoidal beam distribution}
The electrostatic field produced by a uniform ellipsoidal charge distribution of arbitrary aspect ratios is well known in literature. Formally this problem was originally studied in the context of gravitational fields of astrophysical objects. Thus, the literature is mature, and in our analysis we can follow \cite{Kellogg}.

Let us consider a beam traveling in the $z$-direction whose velocity in speed-of-light units is $\beta$. Let us assume that the beam has a charge $Q$ uniformly filling the region enclosed by the ellipsoidal surface with semi-axes $a,b$ and $c^\prime$
\begin{equation}
    \mathcal{E}^\prime: \frac{x^2}{a^2} + \frac{y^2}{b^2} + \frac{{z^\prime}^2}{{c^\prime}^2} = 1
\end{equation}
where all the quantities are expressed in the reference frame co-moving with the beam.
If the energy spread is negligible, the electrons are approximately at rest in such a frame and the
electrostatic potential in a point inside the ellipsoid is given by
\begin{equation}
    \begin{split}
        \Phi^\prime(x,y,z^\prime) &   = \frac{3 Q }{16\pi\epsilon_0}\int_0^{+\infty}\Big( 1 - \frac{x^2}{a^2 + t}+\\
                    &- \frac{y^2}{b^2 + t} - \frac{{z^\prime}^2}{{c^\prime}^2 + t} \Big)\frac{dt}{\sqrt{\varphi(t)}}\\
                    &   \equiv   D_0 - A_0x^2 - B_0y^2 - C_0{z^\prime}^2
    \end{split}
    \label{ellipsoid:potential}
\end{equation}
where $\varphi(t) =   (a^2 + t)(b^2 + t)({c^\prime}^2 + t)$ is an auxiliary function and the coefficients from $A_0$ to $D_0$ depend only on the aspect ratios $a/b$, $a/c^\prime$ and on the total charge.
The exact expressions of such coefficients, in the general case, involve elliptic integrals. However, if the shape of the beam possesses particular symmetries such as the oblate spheroid ($a=b>c^\prime$), the prolate spheroid ($a>b=c^\prime$) and the sphere ($a=b=c^\prime$), simplified expressions can be found \cite{Tuckerman}. 
Inside the distribution, the electrostatic field, $\mathbf{E^\prime} = -\nabla^\prime \Phi^\prime$, derived from the quadratic potential in (\ref{ellipsoid:potential}) grows linearly with the displacement from the center in all three directions yielding linear forces in the beam frame.
In the laboratory frame the linearity of space charge forces is preserved by the Lorentz transformations giving rise to a special dynamical regime: a uniform ellipsoid subjected to linear self-induced forces can expand and contract changing the strength of the forces themselves but it maintains the uniform ellipsoidal shape in the limit of space-charge-dominated behavior. 
As a consequence, in this regime the space-charge forces derived from (\ref{ellipsoid:potential}) provide a self-consistent model for the evolution of the beam, without need to consider emittance-induced dynamics.

The implementation of such an analytical model in a tracking code turns out to be very efficient since it dramatically reduces the number of operations compared to traditional approaches. Indeed, it only requires  calculation of the field coefficients $A_0,B_0$ and $C_0$ from the physical dimensions of the beam.
However, actual beams produced and transported in linacs do not always exhibit nearly ideal ellipsoidal shape. Nonetheless, the formalism shown is still useful as long as the actual distribution is associated to an ``equivalent" uniform ellipsoid, a concept commonly utilized in analyzing and designing beamlines transporting beams with strong space-charge forces \cite{trace3D}. 
In order to accomplish this, we recall that for a uniform ellipsoidal density the laboratory-frame rms dimensions and the beam-frame hard edge semi-axes are related as follows:
\begin{equation}
    \begin{split}
        a & = \sigma_x\sqrt{5}\\
        b & = \sigma_y\sqrt{5}\\
        c^\prime & = \gamma\sigma_z\sqrt{5},
    \end{split}
    \label{eq:hard-edge and std-dev}
\end{equation}
where the relativistic expansion factor appears in the longitudinal direction.
Tracking a beam with arbitrary shape can be accomplished by calculating its rms size and defining, through Eq. (\ref{eq:hard-edge and std-dev}), the semi-axes of an equivalent ellipsoid for which the analytical space charge forces can be applied.
However, it has to be stressed that such a model possesses two main limitations.
On one hand, the transverse wakefield displaces the centroids of the beam slices from the nominal axis. As the model relies on the symmetric shape of the uniform ellipsoid, it fails in case of strong short-range BBU regime which breaks the symmetry of the charge distribution.
Nonetheless, the assumption of quasi-axisymmetric beams for the purpose of calculating space-charge effects remains valid as long as the offset of the slice centroids does not exceed the rms transverse size of the beam or, equivalently, that the transverse emittance growth is small.
If this condition is violated, the analysis is iterated until the assumptions are satisfied, as the design goal of this code is to suppress the BBU and mitigate the emittance growth through suitable correction schemes. Consequently, we will be mainly interested in weak BBU regime which moderately perturbs the symmetry.

In addition, the forces derived from Eq. (\ref{ellipsoid:potential}) do not present a correlation with the longitudinal internal position, or slice.
Thus, effects involving emittance oscillations induced by slices with either different equilibrium-like solutions or a non-homogeneous plasma frequency cannot be described thoroughly.
Examples of such phenomena include the double emittance minimum in a drift space following an rf gun \cite{ferrario2007direct} and the emittance compensation process itself \cite{serafini1997envelope,floettmann2017emittance} for which models exploiting the self-induced fields produced by a set of cylindrical slices \cite{ferrario2007emittance} are often introduced. Thus, we will also present an alternative model based on this concept as a possible improvement in the description of the emittance dynamics.
Nevertheless, as we will show in the following, the method based on ellipsoidal distributions still provides an efficient description of the rms parameters such as the beam envelope and its energy spread. This is particularly true in the main example case we examine in this paper, in which a nearly ellipsoidal beam is produced, and for which a BBU management scheme is extracted which limits the distortions of the ellipsoid due to wakefields and misalignments.

\subsubsection*{Benchmark tests for space charge forces}

In the following we aim to validate the space-charge model based on the equivalent uniform ellipsoid.
Particle beams with different distributions are tracked as they propagate in a long drift space region. Indeed, the absence of applied fields allows investigation of the space-charge effects without the influence of external mechanisms.

For the test we assume monochromatic, psec electron beams with $\SI{250}{pC}$ charge and $\SI{5}{MeV}$ kinetic energy. These are highly space-charge dominated cases. In addition we assume that such beams possess rotational symmetry and that both the transverse and longitudinal phase space have no correlation.

\begin{figure}
    \centering
    \begin{subfigure}{.96\columnwidth}
         \centering
         \includegraphics[width=\textwidth]{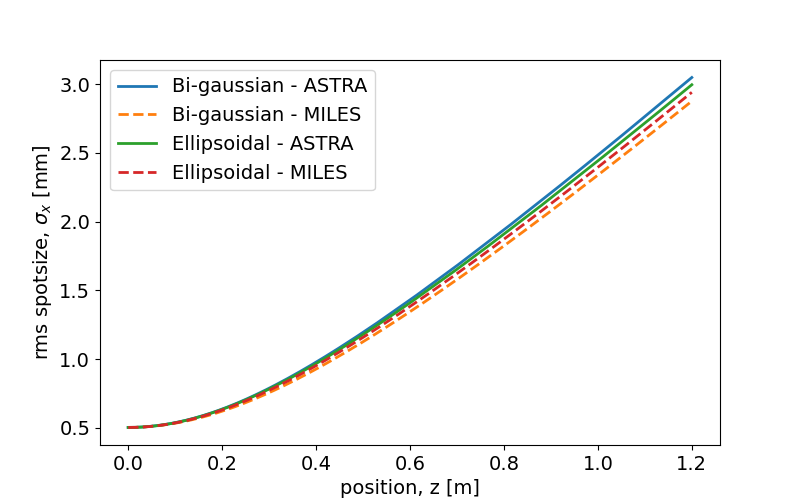}
         \caption{Rms transverse spotsize.}
         \label{std_x_SCon}
     \end{subfigure}
     \hfill
    \begin{subfigure}{.96\columnwidth}
         \centering
         \includegraphics[width=\textwidth]{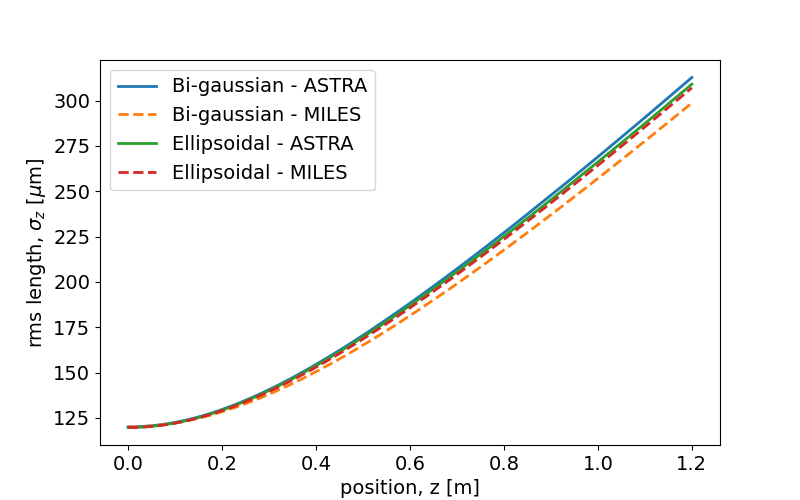}
         \caption{Rms beam length.}
         \label{std_z_SCon}
     \end{subfigure}
     \hfill
     \begin{subfigure}{.96\columnwidth}
         \centering
         \includegraphics[width=\textwidth]{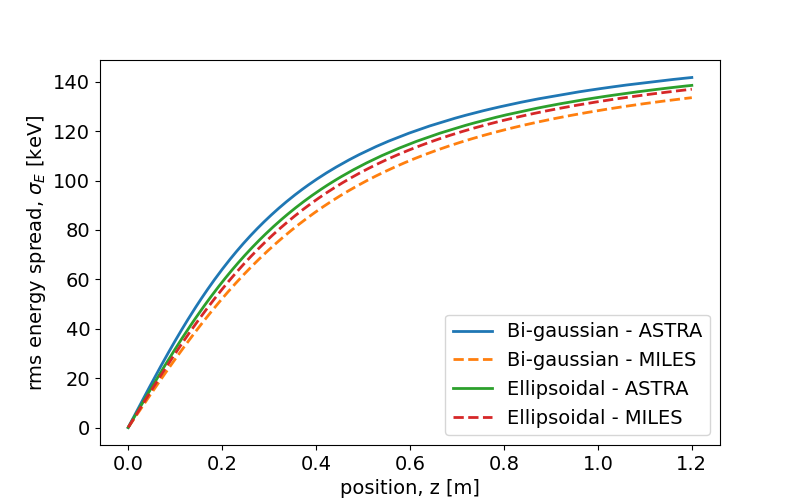}
         \caption{Rms energy spread.}
         \label{std_E_SCon}
     \end{subfigure}
    \caption{Comparison of the rms quantities for different beam distributions propagating in a drift space region.}
    \label{fig:rms_param_SCon}
\end{figure}

The evolution of the rms beam parameters is shown in Fig. \ref{fig:rms_param_SCon} for two different distributions: a bi-gaussian beam (i.e. gaussian either longitudinally and transversely) and a uniform ellipsoid for which particle densities are parabolic in all three directions.
For consistency, the two beams are generated with the same rms size so that $\sigma_x = \sigma_y = \SI{500}{\mu m}$ and $\sigma_z = \SI{120}{\mu m}$.
Comparison of the model adopted by MILES with the results obtained by ASTRA shows a very good agreement for uniform ellipsoidal beams while deviations of the order of a few percent occur for non-ellipsoidal distributions.

\begin{figure}[tbh!]
    \centering
    \begin{subfigure}{.96\columnwidth}
         \centering
         \includegraphics[width=\textwidth]{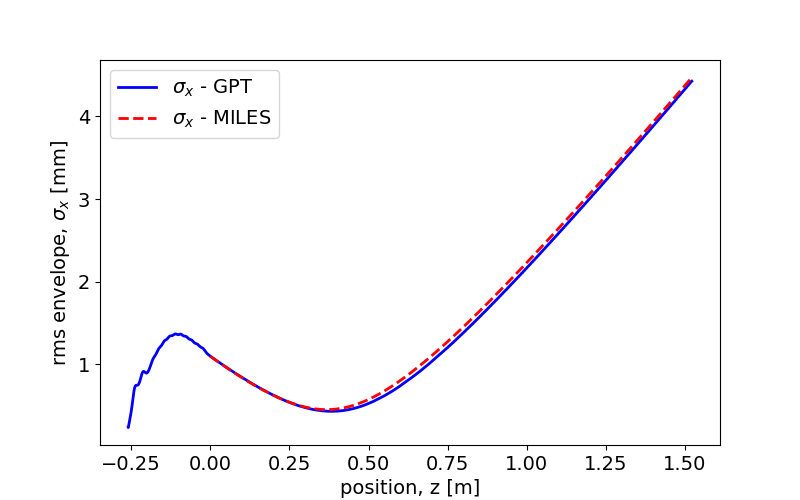}
         \caption{Transverse plane.}
         \label{hybrid_transverse}
     \end{subfigure}
     \hfill
     \begin{subfigure}{.96\columnwidth}
         \centering
         \includegraphics[width=\textwidth]{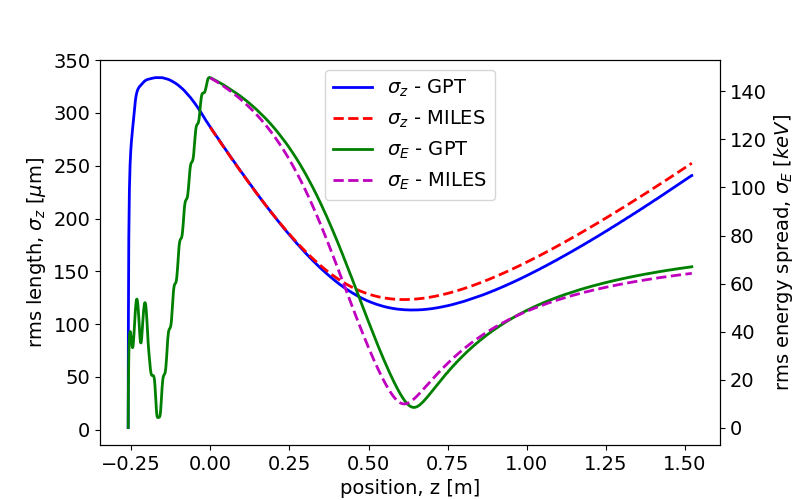}
         \caption{Longitudinal plane.}
         \label{hybrid_longitudinal}
     \end{subfigure}
    \caption{Second order moments of the beam distribution in a $\sim\SI{1.5}{m}$ long drift space.
    Comparison between GPT (solid lines) and custom tracking with ellipsoidal space charge forces (dashed lines). The tracking starts at $z = 0$ where the beam exits the hybrid photoinjector.}
    \label{fig:hybrid_gun_drift}
\end{figure}

An additional example based on a more realistic beam distribution is discussed below.
In Ref. \cite{faillace2021beam,Faillace2022highfield} we presented a C-band hybrid photo-injector producing high brightness electron beams and driving advanced light sources.
Once emitted from the cathode, the beam experiences the typical processes involved in rf guns such as strong rf acceleration, low energy space charge repulsion, focusing through solenoid lenses, chromaticity and, due to the hybrid structure, velocity bunching.
As discussed in the paper, the peculiar combination of such effects results in the formation of quasi-uniform ellipsoidal beams.
Figure \ref{fig:hybrid_gun_drift} shows the evolution of the second order moments in a $\sim\SI{1.5}{m}$ long drift. Solid curves, starting at the cathode, are calculated with the software General Particle Tracer (GPT) \cite{ref_GPT} while dashed lines, starting in $z=0$ at the end of the hybrid gun, are obtained with MILES.
Specifically, the beam coordinates at $z=0$ are exported from GPT and imported in our code to ensure the same initial conditions.
The agreement shown in this comparison illustrates that the ellipsoidal model adopted by MILES is quite satisfactory also for this case concerning a real application.

\begin{figure}
    \centering
    \includegraphics[width=0.96\columnwidth]{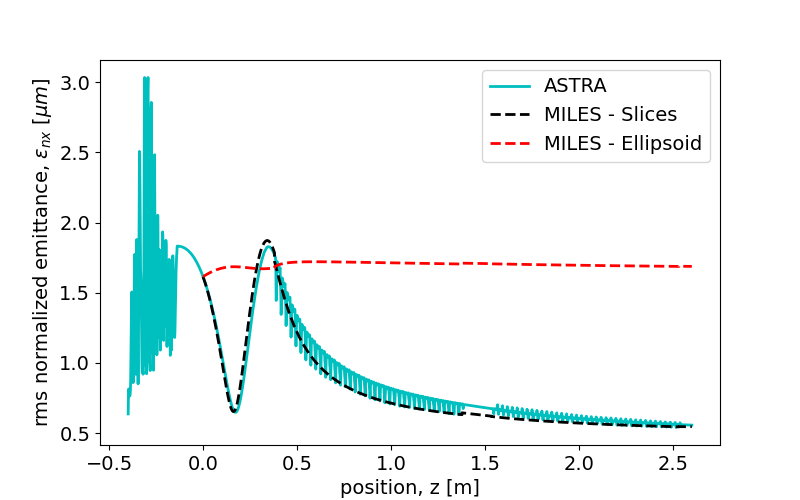}
    \caption{Rms emittance evolution for the beam produced by a C-band hybrid gun and matched to the invariant envelope mode. Comparison among ASTRA (cyan solid line), MILES using 100 cylindrical slices (black dashed line) and MILES using the ellipsoid method (red dashed).}
    \label{fig:emittcompensation}
\end{figure}

As previously mentioned, the equations for ellipsoidal distributions fail to describe observed emittance dynamics effects such as the oscillations in a drift following an rf-gun and the compensation process for beams matched to the invariant envelope.
Before concluding the section we describe an alternative approach based on the use of cylindrical slices which permits description of the evolution of the rms correlated emittance.
This method can be considered a generalization of the model used by the code HOMDYN \cite{ferrario1995Multibunch,Ferrario2000homdyn}.
The latter tracks the evolution of the rms envelope for cylindrical beam slices in a distribution which is near to a uniformly filled cylinder, as well as the motion of their centroids, by exploiting a multi-slice approximation.
Such assumptions successfully describe the dynamics for facilities with designs similar to the standard working point of the LCLS, where $\sim\SI{10}{ps}$ beams are launched with uniform circular cross section and flat-top longitudinal profile. In this scenario, only slight deviations from the original cylindrical, or ``beer-can", shape are expected.
However, such conditions are not found in state-of-the-art facilities producing well sub-ps beams undergoing the process of longitudinal and transverse blowout. This scenario results in the formation (as noted in the reference case for this paper) of quasi-ellipsoidal distributions.
The generalization to modelling beams with more arbitrary shapes that we propose starts by dividing the beam in a finite number of cylindrical slices with individual charge $Q_s$, length $L_s$, radius $R_s$ and mean energy $\gamma_s mc^2$.
In first order, the electric field produced by each slice is provided, either inside and outside the slice itself, by the following compact equations
\begin{subequations}
\label{eq:E_SCfield_cylinder}
\begin{align}
    E_z(0,\zeta_s) & = \frac{Q_s}{2\pi\epsilon_0R_s^2}h(\zeta_s,A_s)\\
    E_x(x_s,\zeta_s) & = \frac{Q_s x_s}{2\pi\epsilon_0 R_s^2 L_s}g(\zeta_s,A_s)
\end{align}
\end{subequations}
where $A_s = R_s/\gamma_s L_s$ is the slice beam-frame aspect ratio, $x_s = x - \langle x\rangle_s$ is the horizontal deviation from the slice axis ($\langle\cdot\rangle_s$ implies an average on the $s$-th slice subset) and $\zeta_s = (z - z_{t,s})/L_s$ accounts for the longitudinal separation from the tail $z_{t,s}$ of the slice.
Moreover, the form functions
\begin{subequations}
\label{eq:hg_SC_cylinder}
\begin{align}
    h(\zeta_s,A_s) & = \vert\zeta_s\vert - \vert 1 - \zeta_s\vert + \sqrt{A_s^2 + (1 - \zeta_s)^2} - \sqrt{A_s^2 + \zeta_s^2}
    \\
    g(\zeta_s,A_s) & = \frac12\left[\frac{1-\zeta_s}{\sqrt{A_s^2 + (1 - \zeta_s)^2}} + \frac{\zeta_s}{\sqrt{A_s^2 +  \zeta_s^2}}\right]
\end{align}
\end{subequations}
are responsible for the correlation between planes as a consequence of the finite length of the bunched beam slice.
A charge $q$ with position $(x,z)$ experiences the force 
\begin{subequations}
    \label{eq:force-slices}
    \begin{align}
        F_x & = q\sum_s E_x(x_s,\zeta_s)/\gamma_s^2\\
        F_z & = q\sum_s E_z(0,\zeta_s)
    \end{align}
\end{subequations}
arising from the superposition of the fields produced by all the slices.
Therefore, such a model allows description of the interplay among different slices which is responsible for the evolution of the correlated emittance.
In addition, Eq. (\ref{eq:force-slices}) is still valid in case the individual slice axes are arbitrarily misaligned from the nominal linac axis.
Thus, the model remains valid in presence of short-range transverse wakefields that displace the centers of the thin cylindrical slices. In Fig. \ref{fig:emittcompensation} we show the rms normalized emittance dynamics for the beam produced by a C-band hybrid photoinjector and injected in a booster linac according to the invariant envelope criterion.
The solid curve is obtained with ASTRA while the black dashed curve is obtained by importing and tracking the beam distribution in our code using 100 cylindrical slices.
The tracking starts at $z = 0$ ($\sim\SI{40}{cm}$ before injection) and allows to reproduce correctly the emittance oscillation in the drift as well as the compensation process. Conversely, the red dashed curve is obtained by using the ellipsoidal model which, due to the absence of correlation, exhibits almost no variations showing the limits of such a model.
It should be stressed that, as the enhanced model for space charge forces relies on the action exerted by all the slices on each macro-particle, the number of operations involved increases notably. As wakefield descriptions are very much amenable to slice-based modelling, the introduction of the enhanced slice model has notable advantages over 3D PIC-based simulations. 
Further, in high brightness photo-injectors followed by booster linacs, variations of the rms emittance are generally relevant only at energies below $\sim$ 100-150 MeV. At higher  energies the beam is no longer space-charge dominated. Therefore, the possibility of ``hybrid" simulations exploiting the cylindrical slice-model at lowest energies, and the ellipsoidal model in the high energy regime can be considered to optimize simulation run-times, while preserving an accurate description of the emittance dynamics. 


\section{Modeling wakefield effects}

The interaction of charged beams with the surrounding environment represents a potentially damaging perturbation of the nominal motion.
The excitation of \emph{wakefields} acting back on the beam itself can cause a significant phase space dilution and needs to be studied thoroughly.
In linacs, the main source of wakefields arises from the accelerating structures. Under conditions that many practical cases (including those examined here) satisfy, the short range interaction for such devices can be evaluated by use of diffraction theory \cite{Lawson,Gluckstern_shape,Heifets:Kheifets}.
The latter allows to find the asymptotic monopole and dipole wake-function per unit length for very short bunches in a periodic cavity whose unit cell has iris dimension $a$, gap size $g$ and period length $p$ \cite{Yokoya:1999ec,Bane:1998hw,Bane2003ShortrangeDW}:
\begin{subequations}
\label{eq:KBane wake}
\begin{align}
    w_\parallel(s) & = \frac{Z_0 c}{\pi a^2}\exp{\left(-\sqrt{\frac{s}{s_0}}\right)}\\
    w_\perp(s) & = \frac{4 Z_0 c s_1}{\pi a^4}\left( 1 - \left( 1 + \sqrt{\frac{s}{s_1}} \right)\exp{\left(-\sqrt{\frac{s}{s_1}}\right)}\right),
\end{align}
\end{subequations}
where $s\geq0$ is the separation from the source particle, $s_0\approx 0.41 a^{1.8}g^{1.6}/p^{2.4}$, $s_1\approx 0.169a^{1.79}g^{0.38}/p^{1.17}$ and $Z_0\approx\SI{377}{\Omega}$ is the impedance of free space.
In addition, the former expressions are valid in the hypothesis of deep cavities ($g\ll 2(b-a)^2/p$, $b$ being the pipe radius) in steady state regime ($z\gtrsim a^2 p /2g\sigma_z$) with $s/p<0.15$, $0.34 \lesssim a/p \lesssim 0.69$ and $0.54 \lesssim g/p \lesssim 0.89$ \cite{Bane2007subps,Palmer1987Aqualitative}.

\subsection{Wakefield matrix formalism}
The wakefield interaction among charged particles can be described in terms of a macro-particle-based matrix formalism which allows evaluation of both longitudinal and transverse effects \cite{Migliorati_MuSiC_1,Migliorati_MuSiC_2}.
Such a method is based on the simple interaction with a resonant cavity mode; it is explained in the following.

A mode with resonant frequency $\omega_0/2\pi$ and quality factor $Q$ has the following longitudinal and dipole wake-functions
\begin{subequations}
    \begin{align}
        w_\parallel(\tau) & = \frac{\omega_0 R_\parallel}{Q}e^{-\alpha\tau}\left(\cos(\omega_n\tau) - \frac{\alpha}{\omega_n}\sin(\omega_n\tau)\right)\\
        w_\perp(\tau) & = \frac{\omega_0 R_\perp}{Q}e^{-\alpha\tau}\sin(\omega_n\tau)
    \end{align}
\end{subequations}
\noindent
where $c\tau\geq0$ is the separation from the source particle, $R_\parallel$ ($R_\perp$) are the longitudinal (transverse) shunt impedances of the mode, $\omega_n = \omega_0\sqrt{1-(2Q)^{-2}}$ and $\alpha = \omega_0/2Q$ \cite{PalumboVaccaroZobov}.
Resonant modes admit a lumped element \textit{RLC}-shunt representation for which the evolution in time of the voltage drop across the capacitance is described by the following differential equation
\begin{equation}
    \ddot{V}(t) + 2\alpha \dot{V}(t) + \omega_0^2 V(t) = 0,
    \label{eq:V differential eq}
\end{equation}
where the dot represents differentiation with respect to time.
Solution of Eq. (\ref{eq:V differential eq}) provides the voltage and its derivative which, together, define the vector $(V, \dot{V})^T$ describing the voltage state of the circuit. The evolution of the system after a time delay $\tau$ can be cast in matrix form:
\begin{widetext}
\begin{subequations}
\label{eq:VVp matrix}
    \begin{align}
        \begin{pmatrix}
            V(t) \\
            \dot{V}(t)
        \end{pmatrix}
        & \mapsto
        \begin{pmatrix}
            V(t+\tau) \\
            \dot{V}(t+\tau)
        \end{pmatrix}
        = 
        M(\tau)
        \begin{pmatrix}
            V(t) \\
            \dot{V}(t)
        \end{pmatrix}\\
        M(\tau) & =
        e^{-\alpha\tau}
        \begin{pmatrix}
            \cos\omega_n\tau + \frac{\alpha}{\omega_n}\sin\omega_n\tau & \frac1{\omega_n}\sin\omega_n\tau\\
            -\frac{\omega_0^2}{\omega_n}\sin\omega_n\tau & \cos\omega_n\tau - \frac{\alpha}{\omega_n}\sin\omega_n\tau
        \end{pmatrix}
    \end{align}
\end{subequations}
\end{widetext}
where the matrix $M(\tau)$, in the context of collective effects, is known as \emph{wakefield matrix}.

Before the beam enters the cavity, the mode is initially unloaded and we find that $(V_0, \dot{V}_0)^T=\mathbf{0}$.
However, the passing beam charges constitute a perturbation of the voltage state, which will evolve accordingly.
For a system receiving such additive perturbations, the linear matrix-transformation described by Eq. (\ref{eq:VVp matrix}) is no longer sufficient to account for its evolution.
The correct transformation operator has the form $\mathcal{T}(\cdot)=\mathbf{p} + M(\cdot)$,
which consists of multiplying the argument to the left by the matrix $M$ and adding the perturbation vector $\mathbf{p}$ afterwards.

For a beam consisting of a sequence of macro-particles with separation $c\tau_1\leq c\tau_2\leq\dots$ from the head, the effect of the resonant mode on the n-th particle depends on the voltage state $(V_n, \dot{V}_n)^T$ after its transit.
The latter is determined by all the perturbations $\mathbf{p_i}$ induced by particles arrived earlier ($i<n$), which will evolve in time.
It can be easily demonstrated that the composition (denoted by the symbol $\otimes$) of $n$ operators with the same form as $\mathcal{T}(\cdot)$ yields the following generalized formula:
\begin{equation}
\begin{split}
    & \mathcal{T}_n\otimes\cdots\otimes\mathcal{T}_{1}(\cdot) = (\mathbf{p_n} + M_n)\otimes\cdots\otimes(\mathbf{p_1} + M_1)(\cdot) = \\
    & = \mathbf{p_n} + \sum_{i=1}^{n-1}\left(\prod_{j=1}^{n-i}M_{n+1-j} \right)\mathbf{p_i} + M_n\cdots M_1(\cdot)\\
    & = \mathbf{p_n} + \sum_{i=1}^{n-1}M(\tau_n - \tau_i)\mathbf{p_i} + M_n\cdots M_1(\cdot),
\end{split}
\label{eq:composition_Top}
\end{equation}
where $M_{k}=M(\tau_{k} - \tau_{k-1}), k=1,2,\dots, n$, and we can arbitrarily set $\tau_0 = 0$. In the last step, the property of the wakefield matrix  $M(\tau_1)M(\tau_2) = M(\tau_2)M(\tau_1) = M(\tau_1 + \tau_2)$ has been used. Equation (\ref{eq:composition_Top}) permits a straightforward interpretation: the last term on the right hand side, being the overall matrix composition, represents the transformation in absence of perturbations while the first two terms account for the evolution of each perturbation through the subsequent time-steps of the system.
In particular, the term accounting for the unperturbed evolution can be dropped since the mode is assumed to be initially unloaded.

Evaluating the voltage states for a set of $N$ macro-particles according to Eq. (\ref{eq:composition_Top}) requires solving $N(N-1)/2$ operations, as much as calculating the wake-potential convolution, which can be time-consuming for large sets.
Nevertheless a clever implementation of this model can improve the efficiency since the $n$-th voltage state is directly connected to the $(n-1)$-th state.
Indeed, a simple manipulation of Eq.~(\ref{eq:composition_Top}) shows that
\begin{equation}
    \begin{pmatrix}
        V_n\\
        \dot{V}_n
    \end{pmatrix}
    =
    \mathbf{p_n} + M(\tau_n - \tau_{n-1})
    \begin{pmatrix}
        V_{n-1}\\
        \dot{V}_{n-1}
    \end{pmatrix}
\label{eq:recursive voltage}
\end{equation}
which reduces the number of operations to $N-1$.
Such a simplification allows for a direct application of the algorithm to the macro-particles rather than the longitudinal slices which, being a physical artifact, introduce an additional cause of numerical noise.

In order to exploit this formalism we should describe the connection between the circuit voltage and the beam dynamics. For longitudinal modes the voltage represents the energy lost per unit charge while for dipole transverse modes it represents a deflecting kick.
In particular the following relations hold:
\begin{subequations}
    \label{eq:momentum kick}
    \begin{align}
        c\Delta p_z & = c \int F_z \,dt = - qq_0 w_\parallel(\tau) \doteq - qV_\parallel(\tau)\\
        c\Delta \mathbf{p_\perp} & = c\int \mathbf{F_\perp} \,dt = qq_0\bm{\rho}_0w_\perp(\tau) \doteq q\bm{V_\perp}(\tau)
    \end{align}
\end{subequations}
which, using the subscript zero for referring to the source particle and letting $\bm{\rho}_0$ be $x_0 \hat{x}$ or $y_0\hat{y}$, allows calculation of the longitudinal and transverse momentum kicks from the voltage.
In a similar way, we can determine the perturbation of the voltage state associated with the transit of a charged particle which excites resonant modes.
Such excitation is given by the wake-functions and their derivatives in the limit $\tau\to0$. Recalling the proportionality with the voltage we obtain
\begin{subequations}
    \begin{align}
        \mathbf{p_i^{(\parallel)}} & = q\frac{\omega_0R_\parallel}{Q}
        \begin{pmatrix}
            1 \\
            -\omega_0/Q
        \end{pmatrix}\\
        \mathbf{p_i^{(\perp)}} & = q\omega_n\frac{\omega_0R_\perp}{Q}x_{i}
        \begin{pmatrix}
            0 \\
            1
        \end{pmatrix},
    \end{align}
\end{subequations}
where $x_{i}$ (or $y_{i}$) is the transverse displacement from the axis of the $i$-th charge causing the excitation of dipole modes.
The latter expressions represent the additive vector in the operator $\mathcal{T}(\cdot)$ and should not be confused with the momentum kicks in Eq. (\ref{eq:momentum kick}).

\begin{figure}
    \centering
    \includegraphics[width=0.96\columnwidth]{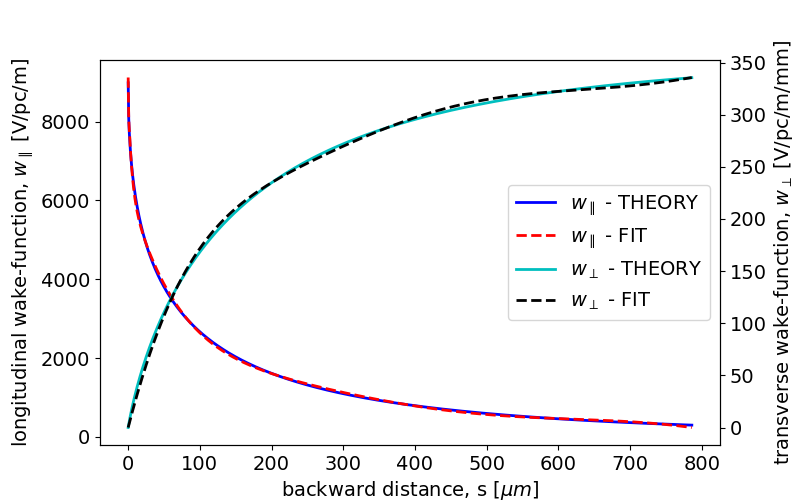}
    \caption{Short-range longitudinal and transverse wake-functions per unit length in a periodic accelerating structures. Solid curves represent formulas from diffraction theory while dashed curves are obtained by a fit with 5 (longitudinal) and 10 (transverse) equivalent resonant modes.}
    \label{fig:fit_wakefunctions_KB}
\end{figure}

So far only the wakefield effects arising from a single resonant mode have been considered. It is straightforward to generalize these results to the case of a finite set of superposed resonant modes. In addition, the formalism allows the representation of arbitrary wake-functions as long as they can be fit by a suitable set of equivalent resonant modes, through the choosing of proper circuit parameters.
As a demonstration for the last statement we consider the well-known short range wake-functions for a periodic accelerating structures in Eq. (\ref{eq:KBane wake}) and we show in Fig. \ref{fig:fit_wakefunctions_KB} the equivalent representation achieved by a fit algorithm \cite{pc_SJ}.

\subsubsection*{Benchmark tests for wakefield effects}

In the remaining part of this section we wish to validate the wakefield matrix formalism described above. In particular, we consider two examples where mono-chromatic rectangular beams interact with a longitudinal broadband resonant mode and with a periodic accelerating structure, respectively.
In both cases it is straightforward to calculate the energy lost by each electron as follows

\begin{equation}
    U_{\text{loss}}(s) = -q Q_bW_\parallel(s) = -q\int w_\parallel (s-s^\prime)\lambda(s^\prime)\,ds^\prime
    \label{eq:Uloss Theory}
\end{equation}
where $\lambda(s), s=c\tau\geq0$ is the longitudinal charge density, $Q_b$ is the beam charge and $W_\parallel(s)$ is the longitudinal wake-potential.

\begin{figure}
    \centering
    \includegraphics[width=0.96\columnwidth]{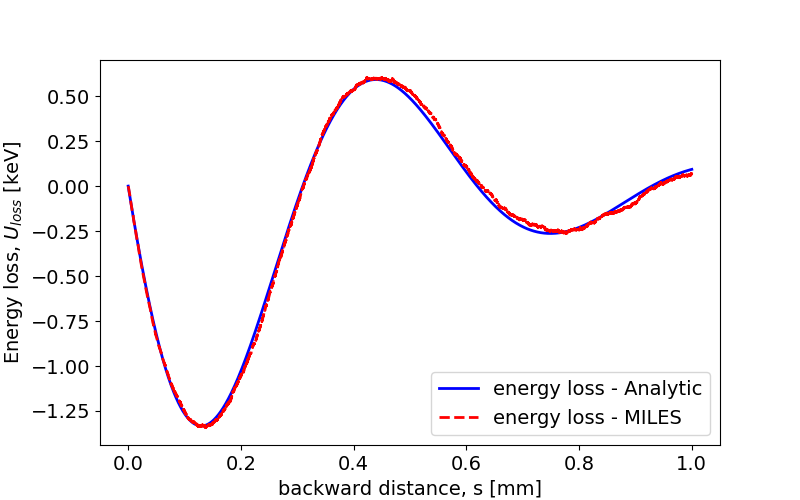}
    \caption{Energy lost by a rectangular beam interacting with a longitudinal broadband HOM.}
    \label{fig:bb_HOM}
\end{figure}

In Fig. \ref{fig:bb_HOM} we show the individual energy loss within a flat $\SI{250}{pC}$ beam of length $l=\SI{1}{mm}$ and kinetic energy $\SI{5}{MeV}$ interacting with a resonant mode whose parameters are $\omega_0/2\pi = \SI{0.5}{THz}$, $Q = 2$ and $R_\parallel = \SI{50}{\Omega}$ in a $\sim\SI{2.62}{cm}$ long cell.
The solid curve corresponds to the energy loss calculated analytically as explained in (\ref{eq:Uloss Theory}) while the dashed curve shows the same quantity calculated by MILES with the matrix formalism presented above.
Note that space-charge forces have been removed in order to focus exclusively on wakefield effects.

\begin{figure}
    \centering
    \includegraphics[width=0.96\columnwidth]{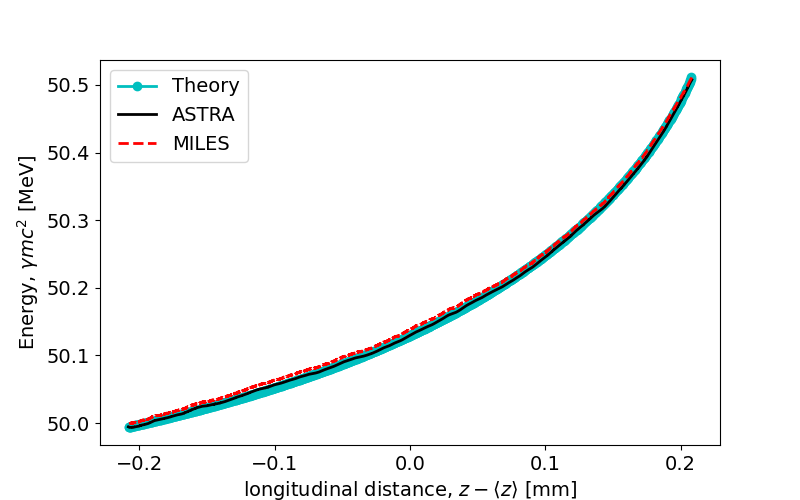}
    \caption{Longitudinal phase space of a rectangular monochromatic beam subjected to longitudinal wakefields in a periodic accelerating structure.}
    \label{fig:KB_longitudinal}
\end{figure}

The next example accounts for the energy lost by a rectangular beam ($Q_b = \SI{250}{pC}$, $l\sim\SI{415}{\mu m}$) injected with a kinetic energy of $\SI{50}{MeV}$ in a periodic accelerating structure.
The linac section in this case is designed using a distributed coupling geometry \cite{tantawi2020design} and is approximately $\sim\SI{1}{m}$ long. The longitudinal wakefield interaction is described by Bane's asymptotic formula for periodic cavities with $\SI{2}{mm}$ iris radii, $\SI{2}{cm}$ cavity gap and $\sim\SI{2.62}{cm}$ cell length.
In order to prevent additional mechanisms from affecting the longitudinal dynamics, space charge forces have been removed again and the rf field in the cavities has zero amplitude to avoid phase dependent acceleration.
In Fig. \ref{fig:KB_longitudinal} the longitudinal phase space downstream the linac section is shown. Specifically, we compare the final energy predicted by Eq. \ref{eq:Uloss Theory} in cyan, the results obtained by tracking with ASTRA in solid black, and with MILES, in dashed red.

\section{Alignment errors}
Linear accelerator sections in real machines are subject to alignment errors of random nature.
Such imperfections introduce deflecting kicks and focusing errors which affect the reference trajectory as well as the beam quality.
Indeed, charged particles moving off-axis in axi-symmetric devices excite dipole transverse wakefields which cause further deflection of the trailing charges, thus reinforcing the excitation process.
The centroid of each slice experiences a transverse displacement due to the wakefields so that a correlation between the planes defined by longitudinal position within the bunch $\zeta$ is introduced. Through this mechanism, the projected rms-emittance grows with a consequent dilution of the phase space.

In the remaining part of this section we will discuss an example in order to show the effects of misalignments in linacs and to verify the model adopted in MILES.
We start by introducing local coordinate systems aimed to describe particles moving inside misaligned linac sections. If we let $(x,z)$ be the transverse and longitudinal coordinates with respect to the nominal machine design trajectory and $(x_i,z_i)$ be the coordinates
with respect to the i-th section's axis, the following transformation holds
\begin{subequations}
    \label{eq:misalignment_transform}
    \begin{align}
        \begin{pmatrix}
            x\\
            z
        \end{pmatrix}
        & =
        \begin{pmatrix}
            \cos\delta_i & \sin\delta_i\\
            - \sin\delta_i & \cos\delta_i
        \end{pmatrix}
        \begin{pmatrix}
            x_i\\
            z_i
        \end{pmatrix}
        +
        \begin{pmatrix}
            {\Delta x}_i\\
            {\Delta z}_i
        \end{pmatrix}\\
        x^\prime & = \frac{x_i^\prime + \tan\delta_i}{1 - x_i^\prime\tan\delta_i},
    \end{align}
\end{subequations}
where $(\Delta x_i, \Delta z_i)$ is the center of i-th structure’s input face
and $\delta_i$ its tilt angle with respect to the nominal axis.
It is understood that, for sections not affected by misalignments, one has $\Delta x_i=0$, $\delta_i=0$ and Eq. (\ref{eq:misalignment_transform}) reduces to the identity with a simple shift in the longitudinal direction.

\begin{figure}[tbh!]
    \centering
    \includegraphics[width=0.96\columnwidth]{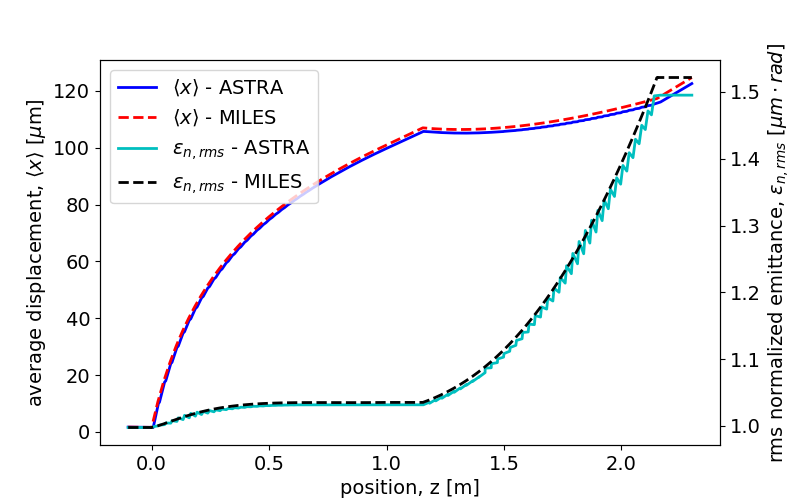}
    \caption{Dipole wakefield interaction in a misaligned two section linac with $\Delta x_1=\SI{100}{\mu m}$ and $\Delta x_2=0$. Comparisons between ASTRA (solid lines) and MILES (dashed lines) are shown for the center-of-mass trajectory and the rms normalized emittance.}
    \label{fig:bm-misaligned section}
\end{figure}

The following example evaluates the dynamics in two sections of a distributed coupling linac where the first one has a transverse offset $\Delta x_1 = \SI{100}{\mu m}$.
Figure \ref{fig:bm-misaligned section} shows a 250 pC, monochromatic electron beam with 5 MeV kinetic energy and rms parameters $\sigma_x = \SI{500}{\mu m}$, $\sigma_z = \SI{120}{\mu m}$ and $\epsilon_{nx} = \SI{1}{\mu m\cdot rad}$ which is injected into and accelerated by the linac.
The alignment error forces the center-of-mass away from the nominal axis, causing excitation of dipole wakefields. Consequently, as the beam propagates inside the linac, the rms emittance grows, as discussed above.
The plot compares the results obtained by ASTRA and MILES, \emph{i.e.} using the matrix formalism, showing good agreement. Notice that space-charge forces have been excluded by both simulations in order to focus exclusively on wakefield effects.


\section{Correction techniques}

Due to the limited precision achievable in fabrication and surveying procedures \cite{alignment}, online correction schemes exploiting beam-based alignment concepts are often employed  \cite{one2one,dispersion_free}.
A proper combination of beam position monitors (BPMs) and dipole magnets with moderate strength can be used to steer the center-of-mass trajectory in order to minimize BBU effects.
In particular, measuring the transverse offset with a set of BPMs allows adjustment of the strength in the steering dipoles accordingly, and provide effective corrections addressing alignment imperfection effects.

In the reminder of this section we discuss two examples of beam-based alignment techniques applied to the first linac stage of an ultra-compact X-ray free-electron laser (UC-XFEL) \cite{ucxfel} based on high-field cryogenic operation of the above-mentioned distributed coupling linac.
In addition, the electron beam is produced by a state-of-the-art C-band cryogenic rf gun \cite{robles2021versatile} operated at $\sim\SI{240}{MV/m}$ peak field.
The main parameters of the beam at the injection into the booster linac are listed in Table \ref{tab:beam_parameters_ucxfel}. The following examples model two $\sim\SI{1}{m}$-long linac sections with accelerating gradient $\SI{77}{MeV/m}$ and $\SI{60}{MeV/m}$ respectively which provide a final energy of approximately $\SI{150}{MeV}$.
It should be mentioned that the working point of such a system foresees compensation of the initial rms emittance (as predicted by \cite{serafini1997envelope}) down to $\sim\SI{55}{nm\cdot rad}$. However, as we have previously stated, the limitations of the ellipsoidal model we adopt for the space charge forces forbid from observing a complete compensation process.

\begin{table}[b]
\caption{\label{tab:beam_parameters_ucxfel}%
Electron beam input parameters for the UC-XFEL \cite{ucxfel}}
\begin{ruledtabular}
\begin{tabular}{lc}
Parameter & Value\\
\hline

Charge, $Q_b$ & \SI{100}{pC} \\ 
           Energy, $\gamma mc^2$ & $\SI{6.9}{MeV}$\\
           Rms spotsize, $\sigma_{x,y}$ & $\SI{60}{\mu m}$ \\ 
           Rms length, $\sigma_z$ & $\SI{0.5}{mm}$ \\ 
           Rms norm. emittance, $\epsilon_{n,\text{rms}}$ & $\SI{100}{nm\cdot rad}$\\
\end{tabular}
\end{ruledtabular}
\end{table}

The first alignment procedure we demonstrate is known as \emph{one-to-one steering} \cite{one2one}. The application of such a method implies an equal number of BPMs and correctors, which we assume to be located downstream and upstream, respectively, of each linac section.
Knowledge of the first order transfer map describing the evolution within the linac allows one to write a linear relation connecting the transverse kick $\Delta x^\prime_j$ applied by the j-th corrector and the transverse offset $X_i$ measured by the i-th BPM. A matrix equation in terms of the integrated magnetic field strengths $K_j$ (in $\SI{}{T\cdot m}$) is obtained
\begin{subequations}
    \begin{align}
        \begin{pmatrix}
        X_1\\
        X_2\\
        \vdots\\
        X_n
    \end{pmatrix}
    & =
    \begin{pmatrix}
        R_{12}^{1\mapsto1} &  &  & \mathbf{0}\\
        R_{12}^{1\mapsto2} & R_{12}^{2\mapsto2} &  & \\
        \vdots & \cdots & 
        \ddots & \\
        R_{12}^{1\mapsto n} & R_{12}^{2\mapsto n} & \cdots & R_{12}^{n\mapsto n}
    \end{pmatrix}
    \begin{pmatrix}
        \Delta x^\prime_1\\
        \Delta x^\prime_2\\
        \vdots\\
        \Delta x^\prime_n
    \end{pmatrix}
    +
    \begin{pmatrix}
        q_1\\
        q_2\\
        \vdots\\
        q_n
    \end{pmatrix}
    \label{one2one:matrix}\\
    \Delta x^\prime_i & = -\frac{qK_i}{\beta_i\gamma_i m c}, \hspace{0.75 cm} i\leq1\leq n
    \end{align}
\end{subequations}
where the vector on the right hand side accounts for the perturbations introduced by the misalignments.
The lower triangular form of the matrix in Eq. (\ref{one2one:matrix}) is due to causality which implies $R_{12}^{i\mapsto j} = 0$ for $j>i$. The physical meaning of such a property is simply that correctors have no influence on measurements performed at BPMs located behind them.
Once the measurements are collected, one-to-one algorithms provide local corrections exploiting Eq. (\ref{one2one:matrix}) and finding the field strength which forces zero transverse offset in the BPMs.

\begin{figure}[h!]
    \centering
    \begin{subfigure}{.96\columnwidth}
         \centering
         \includegraphics[width=\textwidth]{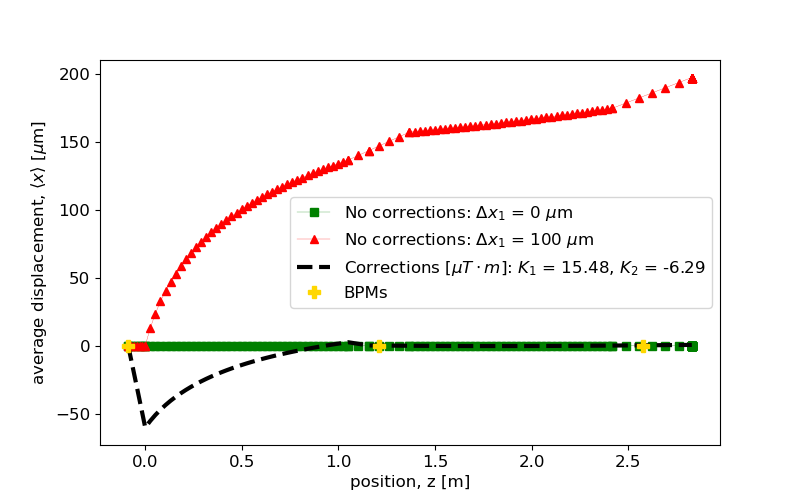}
         \caption{Center of mass trajectory}
         \label{121-center of mass}
     \end{subfigure}
     \hfill
     \begin{subfigure}{.96\columnwidth}
         \centering
         \includegraphics[width=\textwidth]{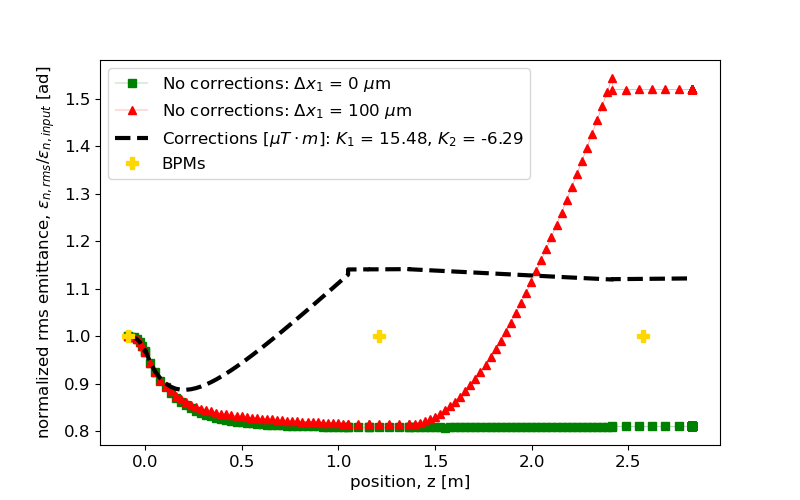}
         \caption{Rms normalized emittance growth}
         \label{121-emittance growth}
     \end{subfigure}
    \caption{Application of one-to-one steering algorithms. The green squares and red triangles represent, respectively, an ideal machine with no misalignments and a linac whose first section is $\SI{100}{\mu m}$ off-axis. Compensation of such an error is achieved by using the optimal corrections provided by the one-to-one procedure as shown by the black dashed curves. The gold crosses identify the location of the BPMs.}
    \label{fig:121-steering}
\end{figure}

Figure \ref{fig:121-steering} shows an application of one-to-one steering assuming that the first linac section exhibits a transverse offset of $\SI{100}{\mu m}$.
In absence of corrections the beam propagates far from the axis of the accelerating structure exciting dipole wakefields.
In this case emittance, normalized by its injection value, grows by up to 50\% as shown by the red triangles.
As a reference, the ideal beam trajectory and emittance evolution for an aligned machine are also shown (green squares).
The black dashed curves represent the result of applying the corrections provided by the one-to-one procedure which, as expected, force the beam trajectory to intersect the BPMs.
As a consequence, the corresponding rms emittance growth is reduced but still present since the algorithm aims for a local optimization of the trajectory rather than a global emittance minimization.
In particular, we notice that the beam is displaced far from the axis while traversing the first linac section, located at $x=\SI{100}{\mu m}$, thus experiencing intense deflections induced by the wakefields.

\begin{figure}[htb!]
    \centering
    \includegraphics[width=0.96\columnwidth]{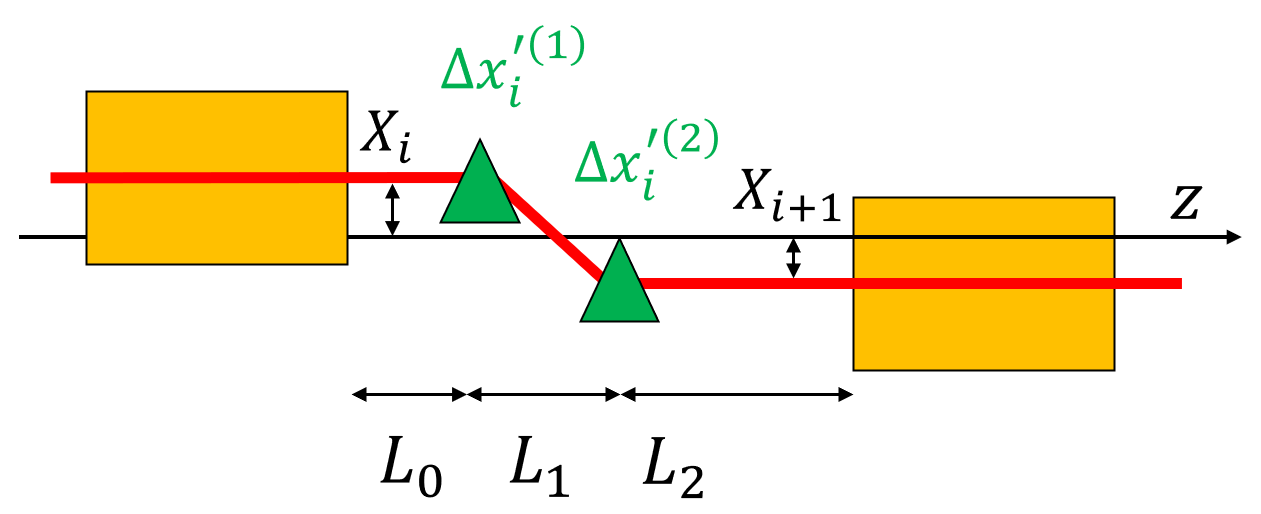}
    \caption{Sketch of the correction scheme matching the center of mass of the beam at the input of each section.}
    \label{fig:sketch wf}
\end{figure}

The previous example suggests that, in order to prevent the emittance from increasing, the excitation of dipole wakefields should be minimized. The fulfillment of such a condition is achieved if the beam is forced to travel on the axis of each accelerating section regardless of the alignment errors.
This fine manipulation of the beam trajectory requires control of both the center-of-mass offset and transverse angle which cannot be achieved with a single magnetic corrector.
However an additional degree of freedom is introduced if two steering dipoles upstream each linac section are used, as schematically shown in Fig. \ref{fig:sketch wf}.
Indeed, any residual error $(X_i,X^\prime_i)$ of the center of mass at the exit of the i-th section can be ideally matched to the input of the (i+1)-th section.
It can be shown that the value of the transverse kicks $\Delta x_i^{\prime(1)}$ and $\Delta x_i^{\prime(2)}$ matching the two sections are given by
\begin{equation}
    \begin{pmatrix}
    \Delta x_i^{\prime(1)}\\
    \Delta x_i^{\prime(2)}
\end{pmatrix}
=
\frac1{L_1}
\begin{pmatrix}
    1 & -L_2\\
    -1 & L_1 + L_2
\end{pmatrix}
\begin{pmatrix}
    X_{i+1} - X_i - L_d X^\prime_i\\
    X^\prime_{i+1} - X^\prime_i
\end{pmatrix}
    \label{eq:matching}
\end{equation}
where $L_d = L_0 + L_1 + L_2$ is the total length of the drift space. Note that, in order to avoid a singular coefficient matrix in the linear system, such an equation strictly requires that $L_1\neq0$.
The physical interpretation of this constraint is that for $L_1\to0$ the two correctors merge into only one and the additional degree of freedom is lost.

\begin{figure}[h!]
    \centering
    \begin{subfigure}{.96\columnwidth}
         \centering
         \includegraphics[width=\textwidth]{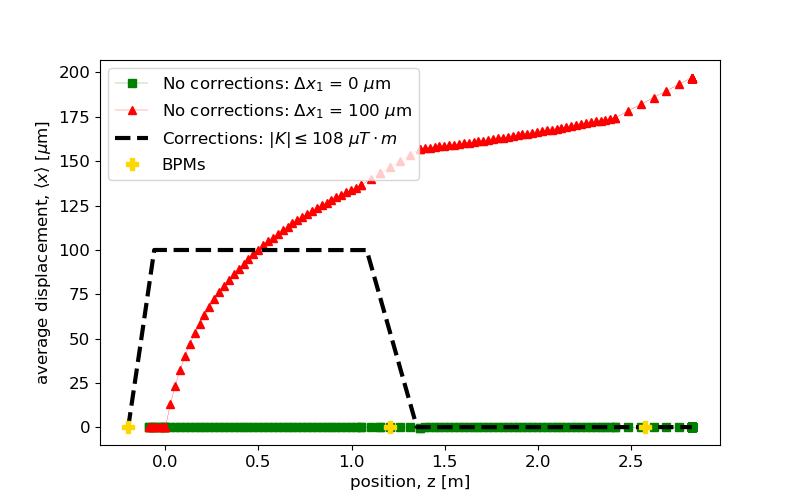}
         \caption{Center of mass trajectory}
         \label{WF-center of mass}
     \end{subfigure}
     \hfill
     \begin{subfigure}{.96\columnwidth}
         \centering
         \includegraphics[width=\textwidth]{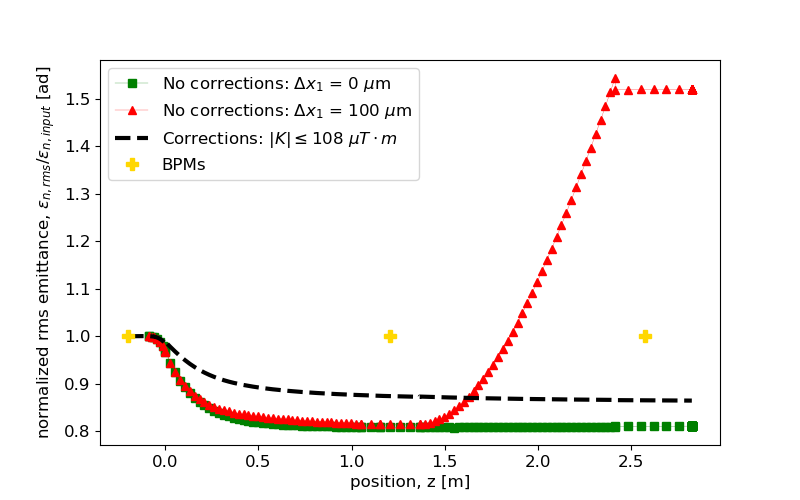}
         \caption{Rms normalized emittance growth}
         \label{WF-emittance growth}
     \end{subfigure}
    \caption{The test described in Fig. \ref{fig:121-steering} is repeated and the same graphic conventions hold. However, the present correction scheme foresees two steering magnets before each linac section which allows to find an optimal trajectory that minimizes the excitation of dipole wakefields.}
    \label{fig:WF-steering}
\end{figure}

To evaluate this scheme, we repeat the same tests performed on the one-to-one steering procedure; we use the same graphics conventions but with a change in the correction scheme that now foresees two steering dipoles per linac section.
In Fig. \ref{fig:WF-steering} the strength of the correctors is optimized to adjust the injection of the beam in each accelerating section.
Consequently, the center-of-mass trajectory always follows the axis of the accelerating structures where the excitation of dipole wakefields is negligible.
Thus, the example shows that the rms emittance can be preserved by applying corrections of reasonable strength such as $\sim 10\div\SI{100}{\mu T\cdot m}$ integrated fields.

As a final remark we want to stress the importance of providing corrections in the first accelerating sections as we have shown in the previous examples.
Indeed, the low energy regime, which is characteristic of photoinjectors followed by a booster linac, causes a high sensitivity to deflections that can rapidly reduce the phase space quality. This sensitivity is exacerbated by the effects of space-charge-dominated operation, which cause the focusing of the beam envelopes and the centroids to behave differently.  
Higher energy beams instead ($\gtrsim\SI{100}{MeV}$) increase their rigidity which mitigates the effect of dipole forces due to either misalignments or transverse wakefields. We note in this regard that the second-order focusing due to RF fields likewise diminishes rapidly with energy. 

\section{Applications}

The examples shown in the previous sections concerning the effects of transverse wakefields induced by alignment errors, and their mitigation, display some of the most relevant applications for which the code MILES was conceived.
Before concluding, we mention a few additional problems of interest that can be investigated as well.

For instance, dedicated studies on the energy spread induced by longitudinal wakefields can be performed. Indeed, such fields are excited even when the beam propagates on-axis negatively affecting the spectral brightness of radiation sources. In addition, compensation techniques adjusting the rf-phase in the accelerating sections could be investigated.

Moreover, the performance of certain types of machines, such as ICS sources or linear colliders, strongly rely on the properties of the beam at the interaction point or IP.
The final focusing of the beam is provided by means of an optical system, for instance a triplet of quadrupole magnets, whose effect can be implemented by means of simple transfer matrices \cite{wolski}.
In case of misalignments a change in the nominal waist position
is expected and, thus, the inclusion of a final focusing system allows one to predict the beam spot size at the actual interaction point.

The matrix formalism presented for short-range wakefields, based on effective resonant modes, additionally allows description of the interaction with the actual higher order modes (HOMs) excited in the linac.
Such modes are responsible for the long-range mutual interaction among different bunches which leads to beam breakup instabilities where the amplitude of the transverse oscillation grows excessively \cite{panofsky1968asymptotic,mosnier1993instabilities}.
Studying these effects becomes extremely important for large scale machines such as TeV-class linear colliders. The ``Cool Copper Collider", or C$^3$ project \cite{Ccube,whitepaperSnowmass}, constitutes a compelling example of this type of machine since it foresees acceleration of intense charged particle beams over the km scale, within the environment of small-iris, distributed coupling linacs.
Therefore, as we have discussed in \cite{COLTRANE}, in addition to the short-range intrabeam effects MILES also allows one to account for the long-range BBU interaction in multi-bunch operation. It additionally includes tools for mitigation techniques based on frequency spread or mode detuning \cite{detuning1,detuning2} breaking the coherent interaction between the resonant modes and the pulsed beam train.




\section{Conclusions}
In this paper we have presented a new tracking code for electron linacs which exploits simple models to account simultaneously for space charge forces and wakefield interaction.
Such features provide a fast, efficient and flexible tool with minimal compromise in the accuracy of the results.  Indeed, comparisons of the results obtained by MILES with well-established models exhibit very good agreement. The main aspects involved in the beam dynamics were successfully validated by means of dedicated benchmark tests whose operating conditions were consistent with the machines of our interest.
We have discussed possible applications that can benefit from the flexibility of this tracking code.
Particular effort was addressed in investigating misalignment effects that induce emittance dilution, as well as possible correction schemes exploiting beam based alignment techniques. In particular we have stressed the importance of providing compensation at low energy, \emph{i.e.} in the first linac sections, where the beam is more sensitive to the imperfections of the machine. With the increasing reliance on creating beams in RF photoinjectors (particularly for light source \cite{Faillace2022highfield,ucxfel} and high energy physics applications\cite{robles2021versatile}) that are directly accelerated in high impedance structures, it is anticipated that the MILES code will be of high use to the beam physics community in the future.


\section{Acknowledgement}
This work is supported by DARPA under Contract N.HR001120C0072, by DOE Contract DE-SC0009914 and DE-SC0020409, by the National Science Foundation Grant N.PHY-1549132 and by the INFN national committee through the project ARYA.


\bibliography{main}



\end{document}